\documentclass[11pt]{article}

\usepackage[margin=1in]{geometry}
\usepackage[T1]{fontenc}
\usepackage[utf8]{inputenc}
\usepackage{lmodern}
\usepackage{float}
\usepackage{amsmath,amssymb,amsthm}
\usepackage{graphicx}
\usepackage{booktabs}
\usepackage{xurl}
\usepackage{natbib}
\usepackage{hyperref}
\usepackage{caption}
\usepackage{authblk}

\graphicspath{{Fig/}}

\hypersetup{
    colorlinks=true,
    linkcolor=blue,
    citecolor=blue,
    urlcolor=blue,
    pdftitle={Physics-Informed Neural Networks for Sparse Strain-Field Reconstruction in 4D-STEM},
    pdfauthor={Roberto dos Reis, Gabriel T. dos Santos, Yukun Liu, Xiaobing Hu, and Vinayak P. Dravid}
}

\title{\textbf{Physics-Informed Neural Networks for Sparse Strain-Field Reconstruction in 4D-STEM}}

\author[1,2]{Roberto dos Reis\thanks{Corresponding author: \href{mailto:roberto.reis@northwestern.edu}{roberto.reis@northwestern.edu}}}
\author[1]{Gabriel T. dos Santos}
\author[1]{Yukun Liu}
\author[2]{Xiaobing Hu}
\author[1,2]{Vinayak P. Dravid}

\affil[1]{Department of Materials Science \& Engineering, Northwestern University, Evanston, IL 60208, USA}
\affil[2]{The NU\textit{ANCE} Center, Northwestern University, Evanston, IL 60208, USA}

\date{}

\begin{document}

\maketitle

\begin{abstract}
Quantitative strain mapping using four-dimensional scanning transmission electron microscopy (4D-STEM) typically relies on densely sampled scans that deliver electron doses high enough to damage many beam-sensitive samples. Here, we develop a physics-informed neural network (PINN) architecture for 4D-STEM strain reconstruction that embeds elastic equilibrium and the Saint-Venant compatibility condition into the training loss through automatic differentiation. The architecture comprises five design choices required by the priors themselves: (i) a coordinate-based implicit \emph{ansatz}, (ii) an activation function with non-vanishing second derivatives, (iii) scale normalization of the physics residuals against the data term, (iv) an exponential physics-weight ramp, and (v) residual-based adaptive collocation. We instantiate the architecture with a sine-activated residual network (SIREN) and apply it to an experimental $180\times400$-pixel strain map of the domain-structured IV--VI high-entropy thermoelectric PbGeSnSe$_{1.5}$Te$_{1.5}$ exhibiting extended chevron-shaped strain bands superposed on pixel-scale measurement noise. Across sampling fractions of $1$--$75\%$ ($720$--$54{,}000$ probe positions, nearly two orders of magnitude in nominal probe-dose budget), the mean absolute error on $\varepsilon_{xx}$ stays within $(1.5$--$2.9)\times10^{-2}$ and $R^2$ reaches $0.80$ at $10\%$ sampling, saturating near $0.86$ by $25\%$; the strain-band morphology is recovered from $10\%$ of the probe positions. At $10\%$ sampling, the physics-informed reconstruction reduces MAE by $\approx26\%$ relative to compressed sensing and $\approx22\%$ relative to Gaussian-process regression. An ablation against a data-only SIREN of identical capacity shows that the PDE prior improves accuracy under extreme sparsity ($1\%$ sampling) and consistently improves physical self-consistency, but becomes a bias when data are abundant. Bayesian approaches, including Monte Carlo dropout and mean-field variational inference, provide per-pixel epistemic uncertainty maps whose values are correlated with true reconstruction error. With a constitutive model appropriate to the specimen, the framework can be adapted to strain mapping in semiconductors, oxides, 2D materials, and battery and thermoelectric phases, among other material systems.
\end{abstract}


%======================================================================
\section{Introduction}
%======================================================================

Four-dimensional scanning transmission electron microscopy (4D-STEM) records a full convergent-beam electron diffraction (CBED) pattern at every position of a raster-scanned probe, producing a dataset indexed by two real-space and two reciprocal-space coordinates~\citep{Ophus2019}. Bragg-disk fitting in each CBED pattern recovers local diffraction-vector shifts relative to a reference lattice, from which the 2D strain tensor $\varepsilon_{ij}(x,y)$ and the derived principal-strain orientation $\theta(x,y)$ are obtained at picometer-level precision~\citep{Pekin2017,Padgett2020,Muller2012}. Open-source pipelines such as py4DSTEM~\citep{Savitzky2021} and pyXEM~\citep{pyxemorientationmapping2022} have made the data analysis workflow widely accessible and the technique is now used routinely for nanoscale strain measurement in semiconductor channels, heteroepitaxial films, ferroelectric oxides, 2D materials, and functional phases for energy storage and thermoelectrics. However, the dense 4D-STEM acquisition matrices required to obtain reliable strain maps can be problematic because they deposit cumulative electron doses that exceed the damage threshold for many technologically relevant specimens. For example, organic semiconductors, hybrid perovskites, metal--organic frameworks, battery cathodes, and several thermoelectric compositions undergo beam-induced segregation, amorphization or cation migration well before a conventional strain map is complete~\citep{Egerton2019}. 

Dense acquisition is also challenging for \textit{in situ} and operando 4D-STEM. Time-resolved experiments further multiply this data burden across successive frames, increasing demands on detector throughput, storage, transfer, and analysis~\citep{spurgeon2021towards}. In addition, the specimen can change over the course of a scan, so that extended acquisition times smear out or merge short-lived structural states. A common strategy is therefore to undersample the probe grid, reducing cumulative electron dose, acquisition time, and data volume. This, however, leads to an ill-conditioned reconstruction task, in which a continuous, vector-valued strain field has to be inferred from sparse and noisy observations that are available only at pixels where Bragg-disk fitting is applicable.

Standard sparse-reconstruction approaches only partially address this. Bilinear and bicubic interpolation impose no physical constraints and oscillate near high-gradient features such as dislocation cores and inclusion boundaries~\citep{Press2007}. Gaussian-process regression provides Bayesian uncertainty estimates but scales as $\mathcal{O}(N^3)$ in the number of observations and admits no native PDE prior~\citep{Rasmussen2006}. Compressed sensing exploits sparsity in a chosen basis (Fourier, wavelet, curvelet) and has demonstrated 2--5$\times$ dose reductions in HAADF and BF STEM~\citep{Kovarik2013,Stevens2018,Candes2006,Donoho2006}, but does not enforce elastic equilibrium or Saint-Venant compatibility. Convolutional and generative networks learn powerful image priors~\citep{Madsen2018,Wang2020,Ziatdinov2021,Reichstein2019,Munshi2022}, yet they require large labeled training sets that rarely exist for novel compositions, and their reconstructions are free to violate the governing PDEs of continuum mechanics.

\subsection{Physics-informed neural networks}

Physics-informed neural networks (PINNs)~\citep{Raissi2019,Karniadakis2021,Cuomo2022,Lu2021} replace the data prior of classical deep learning with an equation prior. The network is queried at a dense set of collocation points; the residual of a governing PDE is evaluated at each point through automatic differentiation and added to the loss. The result is a differentiable, mesh-free representation of the field that is data-consistent at sampled locations and regularized toward satisfaction of the governing equations throughout the domain. PINNs have produced quantitative agreement with reference solutions in fluid, solid, and quantum mechanics~\citep{Raissi2020,Cai2021,Carleo2019,Mao2020,Goswami2020,Haghighat2021,Henkes2022,Rao2021}, often with $10$--$100\times$ fewer samples than purely data-driven baselines.

For 4D-STEM strain analysis, Saint-Venant compatibility is a broadly applicable kinematic constraint for sufficiently smooth, displacement-derived strain fields, whereas elastic equilibrium requires a constitutive closure appropriate to the specimen. In this work, homogeneous isotropic linear elasticity is used as a useful approximate prior. Both expressions involve second spatial derivatives of the strain. Therefore, the network \emph{ansatz} has to support stable, non-saturating second-order automatic differentiation. Sine-activated coordinate networks (SIREN)~\citep{Sitzmann2020} and Fourier-feature networks~\citep{Tancik2020} are two architectures with this property; we use a SIREN backbone in the worked example below.

In this work, we present a PINN architecture for 4D-STEM strain reconstruction in which the architectural choices are driven by the priors. The main contributions are:
\begin{enumerate}
    \item A general architecture that fixes the five key components required by the elastic-equilibrium and Saint-Venant priors: a coordinate-based implicit \emph{ansatz}, an activation with non-vanishing higher derivatives, scale normalization of the physics residuals, a physics-weighted composite loss with an exponential ramp, and residual-based adaptive collocation refinement. We demonstrate that the normalization must be \emph{frozen} after a brief warmup period; if the residual is continually re-normalized, it exerts a pull of constant magnitude that is effectively equivalent to minimizing $\log\mathcal{L}$, and over long optimization runs this ultimately drives the system toward the trivial, strain-free solution.
    \item A concrete instantiation with a SIREN backbone trained on an experimental 4D-STEM highly complex strain map of the IV--VI high-entropy thermoelectric PbGeSnSe$_{1.5}$Te$_{1.5}$~\citep{liu2024} ($180\times400 = 72{,}000$ pixels). We quantify reconstruction quality across sampling fractions spanning two orders of magnitude ($1\%$ up to $75\%$), against compressed-sensing and Gaussian-process baselines evaluated on the same sampling masks.
    \item An ablation against a SIREN of identical architecture with $\lambda_{\text{phys}}=0$, isolating the value of the PDE prior from network capacity. The prior improves accuracy only in the extreme-sparsity regime ($\approx 10\%$ lower MAE at $1\%$ sampling), is neutral near $5\%$, and biases the fit once data are plentiful, while always improving physical self-consistency (equilibrium residuals $1.5$-$1.8\times$ lower, compatibility residuals $2.2$-$3.9\times$ lower). We argue this is the physically expected behavior for a domain-structured crystal, where the spontaneous (transformation) eigenstrain of the ferroelastic domains makes homogeneous elastic equilibrium only an approximate prior on the measured strain.
    \item Two Bayesian variants of the same architecture (Monte Carlo dropout~\citep{GalGhahramani2016} and mean-field variational inference~\citep{Blundell2015,Yang2021BPINN}) that produce per-pixel epistemic uncertainty maps whose magnitude correlates with the actual reconstruction error, together with the stochastic-layer design rules (dropout placement, prior scale) that sine networks require for these schemes to train at all.
\end{enumerate}
The result is an acquisition-and-reconstruction protocol that returns a denoised strain tensor regularized toward the imposed mechanical constraints from $\sim 10\%$ of the probe positions of a conventional dense scan, with a spatially resolved uncertainty map attached to every reconstructed pixel.

%======================================================================
\section{PINN architecture for 4D-STEM strain reconstruction}
\label{sec:arch}
%======================================================================

This section develops the architecture independently of any specific specimen and any specific network family. We (i) state the inverse problem and identify the three sources of ill-posedness it carries, (ii) derive the priors, (iii) read the architectural requirements off the priors, and (iv) define the loss and the training protocol that meet those requirements.

%----------------------------------------------------------------------
\subsection{The 4D-STEM strain inverse problem}
\label{sec:inverse}
%----------------------------------------------------------------------
Consider a thin specimen occupying a 2D domain $\Omega\subset\mathbb{R}^2$ in the plane perpendicular to the electron beam. At every probe position $\mathbf{x}=(x,y)\in\Omega$, Bragg-disk fitting of the local CBED pattern returns local diffraction-vector information (equivalently, local reciprocal-lattice geometry) relative to a chosen reference lattice. From these fitted diffraction vectors one estimates the in-plane deformation gradient and, after symmetrization, the in-plane strain tensor components and a derived principal-strain orientation. In this work we take the resulting derived field to be:

\begin{equation}
\mathbf{u}(\mathbf{x})\;=\;\bigl[\,\varepsilon_{xx}(\mathbf{x}),\,\varepsilon_{yy}(\mathbf{x}),\,\varepsilon_{xy}(\mathbf{x}),\,\theta(\mathbf{x})\,\bigr]^\top \in \mathbb{R}^4 ,
\label{eq:strain-vector}
\end{equation}
where the strain components are defined relative to the reference lattice parameter (or reference reciprocal-lattice vectors) used in the Bragg-disk analysis.

A conventional 4D-STEM acquisition produces $\mathbf{u}$ at every pixel of a dense raster $\Omega_h\subset\Omega$ with spacing $h$ set by the desired spatial resolution. Under dose-limited conditions we observe $\mathbf{u}$ only on a sparse subset $\mathcal{S}\subset\Omega_h$, and only at pixels where Bragg-disk fitting succeeded, indicated by a binary mask $\mathcal{M}(\mathbf{x})\in\{0,1\}$. The measurement model is
\begin{equation}
\mathbf{u}_i^{\text{meas}} \;=\; \mathbf{u}(\mathbf{x}_i) + \boldsymbol{\eta}_i ,\qquad \mathbf{x}_i\in\mathcal{S},
\label{eq:measurement}
\end{equation}
with $\boldsymbol{\eta}_i$ an additive (non-stationary, generally non-Gaussian) noise term reflecting limited electron statistics, pattern overlap and disk-fitting residuals. The reconstruction task is to recover a continuous field $\hat{\mathbf{u}}:\Omega\to\mathbb{R}^4$ from $\{(\mathbf{x}_i,\mathbf{u}_i^{\text{meas}})\}_{i\in\mathcal{S}}$.

The problem is ill-posed in the Hadamard sense on three independent axes. First, \emph{underdetermination:} for $|\mathcal{S}|\ll|\Omega_h|$ the data constrain $\hat{\mathbf{u}}$ only on $\mathcal{S}$, and an infinite-dimensional family of interpolants fits the samples between them. Second, \emph{noise amplification:} differential quantities such as rotation, shear gradient and displacement scale as derivatives of $\hat{\mathbf{u}}$ and amplify per-pixel noise by factors proportional to $h^{-k}$ for the $k$-th derivative. Third, \emph{channel coupling:} the four components of $\mathbf{u}$ are not independent in any physical specimen; they are coupled through equilibrium and compatibility, but channel-wise interpolation discards this coupling. A successful reconstruction algorithm must interpolate, denoise, and enforce inter-channel coupling at the same time. Classical sparse-reconstruction tools address at most two of these three; PINNs address all three in a single differentiable optimization.

%----------------------------------------------------------------------
\subsection{Physics priors for 2D strain}
\label{sec:priors}
%----------------------------------------------------------------------
Two continuum-mechanics constraints regularize the ill-posedness discussed in Section~\ref{sec:inverse}. Both constraints depend on derivatives of the strain field; compatibility is kinematic, whereas equilibrium additionally depends on the adopted constitutive closure.

\subsubsection{Elastic equilibrium}
Under quasi-static conditions and in the absence of body forces, the Cauchy stress tensor $\boldsymbol{\sigma}$ satisfies
\begin{equation}
\nabla\!\cdot\!\boldsymbol{\sigma}\;=\;\mathbf{0}
\quad\Longleftrightarrow\quad
\partial_x\sigma_{xx}+\partial_y\sigma_{xy}=0,\;\;
\partial_x\sigma_{xy}+\partial_y\sigma_{yy}=0 .
\label{eq:equilibrium}
\end{equation}
For an isotropic linear-elastic material with Lam\'e parameters $(\lambda,\mu)$ the constitutive law closes the system,
\begin{equation}
\label{eq:hooke}
\begin{aligned}
\sigma_{xx} &= \lambda(\varepsilon_{xx}+\varepsilon_{yy}) + 2\mu\varepsilon_{xx},\\
\sigma_{yy} &= \lambda(\varepsilon_{xx}+\varepsilon_{yy}) + 2\mu\varepsilon_{yy},\\
\sigma_{xy} &= 2\mu\varepsilon_{xy}.
\end{aligned}
\end{equation}
For the plane-stress implementation used here, the effective elastic constants are
\[
\lambda_{\mathrm{ps}}=\frac{E\nu}{1-\nu^2},
\qquad
\mu=\frac{E}{2(1+\nu)}.
\]
yielding two coupled second-order linear PDEs on $\varepsilon_{ij}$ that every admissible strain field must fulfill at every point. Equilibrium is material-parameter-dependent but geometry-independent. In compositionally graded or multi-phase specimens, $(\lambda,\mu)$ become spatial fields that can be supplied from independent measurement or learned jointly with the network parameters~\citep{Tartakovsky2020,Henkes2022}.

\subsubsection{Saint-Venant compatibility}
Any strain field derived from a continuous displacement $\mathbf{u}_d:\Omega\to\mathbb{R}^2$ via $\varepsilon_{ij}=\tfrac{1}{2}(\partial_j u_{d,i}+\partial_i u_{d,j})$ must satisfy the Saint-Venant compatibility condition. In 2D this collapses to a single scalar equation:
\begin{equation}
\partial_y^2\varepsilon_{xx} \;+\; \partial_x^2\varepsilon_{yy} \;-\; 2\,\partial_x\partial_y\varepsilon_{xy} \;=\; 0 .
\label{eq:compat}
\end{equation}
Compatibility is independent of material parameters. In the absence of eigenstrain, plasticity, dislocations, phase transformations, and other incompatible sources, violation of this condition indicates that the strain field cannot be generated by a single-valued continuous displacement field.

\subsubsection{Effect of priors on the inverse problem}
Together, Eqs.~(\ref{eq:equilibrium})-(\ref{eq:compat}) reduce the effective dimensionality of the admissible-field manifold from $\mathcal{O}(|\Omega_h|)$ (one unconstrained vector per pixel) to a much smaller manifold parameterized, locally, by a 2D displacement field that solves the coupled linear elastostatic problem. The priors project the \emph{ansatz} space onto the image of the displacement-to-strain map, and the data term selects the correct representative within that image. The number of samples needed for accurate recovery scales with the essential dimensionality of the displacement, not with $|\Omega_h|$. This is the structural reason, independent of any specific specimen, that physics-informed reconstruction outperforms data-only reconstruction in the sparse regime~\citep{Karniadakis2021,Cuomo2022}.

%----------------------------------------------------------------------
\subsection{Architectural requirements imposed by the priors}
\label{sec:requirements}
%----------------------------------------------------------------------
Eqs.~(\ref{eq:equilibrium})--(\ref{eq:compat}) impose five specific requirements on any network \emph{ansatz} $f_{\boldsymbol{\phi}}$ used to parameterise $\hat{\mathbf{u}}$. Each requirement maps onto a concrete architectural choice; together, these five choices define the PINN architecture for 4D-STEM strain used here.

\paragraph{Requirement 1: a coordinate-based implicit \emph{ansatz}.}
The priors must be evaluated at arbitrary $\mathbf{x}\in\Omega$, not only on the acquisition grid $\Omega_h$, because the samples in $\mathcal{S}$ in general do not respect any regular spacing. A discretized representation (a CNN over $\Omega_h$, a U-Net, a transformer over patches) cannot evaluate the PDE residual outside its grid. We therefore parameterize $\hat{\mathbf{u}}$ as an implicit neural field
\begin{equation}
\hat{\mathbf{u}}(\mathbf{x}) = f_{\boldsymbol{\phi}}(\mathbf{x})
= \bigl[
\hat{\varepsilon}_{xx},
\hat{\varepsilon}_{yy},
\hat{\varepsilon}_{xy},
\hat{\theta}
\bigr]^\top(\mathbf{x})
\label{eq:ansatz}
\end{equation}

a fully connected network mapping a continuous coordinate to the three strain components and the derived orientation channel, with shared parameters $\boldsymbol{\phi}$. The \emph{ansatz} is meshless, smoothly differentiable, and globally parameterized, so any inductive bias introduced at one point in $\Omega$ regularizes the prediction throughout.

\paragraph{Requirement 2: non-vanishing second derivatives.}
Compatibility (Eq.~\ref{eq:compat}) is second-order in the strain. ReLU networks have piecewise-zero second derivatives, while saturating activations (tanh, sigmoid) have second derivatives that vanish over wide regions of input space~\citep{Krishnapriyan2021}. Both produce ill-posed compatibility losses. The activation function must be smooth and non-saturating to second order. Two architectures meet this requirement: sine activations (SIREN)~\citep{Sitzmann2020}, with $f''=-\omega_0^2 f$ and therefore non-vanishing throughout the domain, and Fourier-feature networks~\citep{Tancik2020}, which embed coordinates through a sinusoidal feature map. We use the SIREN family in this work; the Fourier-feature variant is an equivalent choice.

\paragraph{Requirement 3: residual scale normalization, frozen after warmup.}
The data term is an $\mathcal{O}(1)$ MSE on standardized fields, whereas the physics residuals involve first and second spatial derivatives of those fields over normalized coordinates and are $\mathcal{O}(10^2)$--$\mathcal{O}(10^4)$ at typical feature scales. With fixed weights on the order of 0.1, the physics gradient overwhelms the data gradient, causing the optimizer to settle on the trivial solution of the PDE residuals, namely a spatially uniform field with no strain. We resolve this by dividing each residual by a running estimate of its own magnitude (an exponential moving average accumulated over the first $200$ optimizer steps) and then \emph{freezing} that estimate. The freeze is essential because a continuously updated normalizer keeps the physics pull at a constant magnitude no matter how small the residual becomes, which is equivalent to minimizing $\log\mathcal{L}$, and once the data term plateaus it gradually steers any sufficiently long optimization towards the strain free solution. A frozen normalizer instead turns the physics term into a fixed-scale penalty whose pull decays as the residual shrinks.

\paragraph{Requirement 4: a physics-weight ramp.}
Even correctly normalized, physics gradients dominate early in training, when $f_{\boldsymbol{\phi}}$ is far from any admissible solution, and stall optimization~\citep{Wang2021Pathology,Wang2022}. We therefore additionally ramp the physics weight exponentially in time (Section~\ref{sec:loss}). This is standard PINN practice and is critical when the priors are second-order, because second-derivative gradients have larger condition number than first-derivative gradients.

\paragraph{Requirement 5: adaptive collocation refinement.}
Uniformly sampled collocation points waste computation on regions where the PDE is already satisfied. Residual-based adaptive refinement (RAR)~\citep{McClenny2020,Wu2023} concentrates collocation density on regions of large physics residual (boundaries, high-gradient features). Because the compatibility residual contains second spatial derivatives of strain, high-gradient features and their boundaries can carry disproportionately large residuals and are systematically underrepresented by uniform collocation on $\Omega$. As a result, uniform-collocation PINNs leave significant residual mass on feature boundaries.

We now make each of these five choices concrete in the next two subsections.

%----------------------------------------------------------------------
\subsection{Loss formulation and automatic differentiation}
\label{sec:loss}
%----------------------------------------------------------------------
\subsubsection{Data term}
With $\mathcal{S}$ the set of sampled probe positions and $\mathcal{M}_i\in\{0,1\}$ the valid-pixel mask, the data loss is a masked MSE:
\begin{equation}
\mathcal{L}_{\text{data}}(\boldsymbol{\phi}) \;=\; \frac{1}{|\mathcal{S}|}\sum_{i\in\mathcal{S}}\mathcal{M}_i\;\bigl\|f_{\boldsymbol{\phi}}(\mathbf{x}_i)-\mathbf{u}_i^{\text{meas}}\bigr\|_2^2 .
\label{eq:ldata}
\end{equation}

\subsubsection{Physics terms via autodiff}
Let $\mathcal{C}=\{\mathbf{x}_j\}_{j=1}^{N_c}\subset\Omega$ be a set of $N_c$ collocation points, resampled uniformly at random at every training step. The equilibrium and compatibility residuals are
\begin{align}
\mathcal{L}_{\text{equil}}(\boldsymbol{\phi}) &= \frac{1}{N_c}\sum_{j=1}^{N_c}\Bigl[\bigl(\partial_x\sigma_{xx}+\partial_y\sigma_{xy}\bigr)^{2}+\bigl(\partial_x\sigma_{xy}+\partial_y\sigma_{yy}\bigr)^{2}\Bigr]_{\mathbf{x}_j}, \label{eq:lequil}\\[2pt]
\mathcal{L}_{\text{compat}}(\boldsymbol{\phi}) &= \frac{1}{N_c}\sum_{j=1}^{N_c}\bigl(\partial_y^2\varepsilon_{xx}+\partial_x^2\varepsilon_{yy}-2\,\partial_x\partial_y\varepsilon_{xy}\bigr)^{2}_{\mathbf{x}_j}, \label{eq:lcompat}
\end{align}

with all strains and stresses evaluated through $f_{\boldsymbol{\phi}}(\mathbf{x}_j)$ and Eq.~(\ref{eq:hooke}). The spatial derivatives are obtained exactly by two passes of reverse-mode automatic differentiation of $f_{\boldsymbol{\phi}}$ with respect to its input coordinate~\citep{Paszke2019}. No finite-difference stencil is used, no discretization error enters, and the derivatives are available at arbitrary $\mathbf{x}\in\Omega$. This constitutes the mechanical benefit of adopting the implicit neural field \emph{ansatz} instead of a discretized representation. The physics loss costs $\mathcal{O}(N_c)$ per optimizer step and parallelizes on a single GPU; we use $N_c=2{,}000$ fresh points per step, augmented by the persistent RAR pool.

\subsubsection{Composite objective}
The total loss combines the data term with \emph{scale-normalized} physics residuals under an exponential weight ramp:
\begin{equation}
\label{eq:total}
\begin{aligned}
\mathcal{L}_{\text{total}}(\boldsymbol{\phi},t)
&= \lambda_{\text{data}}\mathcal{L}_{\text{data}}
+ \lambda_{\text{phys}}(t)\Bigl[
\alpha_{\text{eq}}\frac{\mathcal{L}_{\text{equil}}}{s_{\text{eq}}}
+ \alpha_{\text{co}}\frac{\mathcal{L}_{\text{compat}}}{s_{\text{co}}}
\Bigr], \\
\lambda_{\text{phys}}(t)
&= \lambda_{\text{phys}}^{\max}\bigl(1-e^{-t/\tau}\bigr).
\end{aligned}
\end{equation}

where $s_{\text{eq}},s_{\text{co}}$ are exponential moving averages ($\beta=0.99$) of the respective residuals, accumulated over the first $t_{\text{w}}=200$ optimizer steps and frozen thereafter (Requirement~3). Hyperparameters are $\lambda_{\text{data}}=1.0$, $\alpha_{\text{eq}}=\alpha_{\text{co}}=0.1$, $\lambda_{\text{phys}}^{\max}=1.0$, $\tau=500$\,epochs. The normalization makes the physics gradients commensurate with the data gradient regardless of the raw residual magnitude; the freeze prevents the log-loss collapse described in Section~\ref{sec:requirements}; and the ramp lets the network fit the data first and then progressively enforces the PDE constraints, stabilizing training when the priors are second-order~\citep{Wang2022}.

%----------------------------------------------------------------------
\subsection{Training protocol, adaptive refinement and Bayesian extensions}
\label{sec:training}
%----------------------------------------------------------------------
Training uses Adam~\citep{Kingma2015} with initial learning rate $10^{-3}$, exponential decay $\gamma=0.95$ every $500$ epochs, global-norm gradient clipping at $1.0$, and a maximum of $5{,}000$ epochs. Early stopping monitors validation loss with patience $300$, and the checkpoint with the lowest validation loss is restored before reporting reconstructions and evaluation metrics.

\paragraph{Residual-based adaptive refinement.}
From epoch $2{,}000$ onwards, every $500$ epochs we score a fresh pool of $5{,}000$ candidate points by their equilibrium residual and append the top $10\%$ of $N_c$ (i.e., $200$ points) to a persistent collocation pool, capped at $5\times N_c$~\citep{McClenny2020,Wu2023}. This procedure constitutes the PINN-based analogue of adaptive mesh refinement, where computational resources are preferentially allocated to regions in which the governing partial differential equations exhibit the largest residuals, i.e., are most difficult to satisfy, at the expense of an $\mathcal{O}(1)$ increase in the total number of collocation points.

\paragraph{Bayesian variants.}
A point estimate $f_{\hat{\boldsymbol{\phi}}}$ from a single run does not quantify the epistemic uncertainty induced by sparse data and a non-convex landscape. Two complementary posterior schemes apply to the architecture above, but sine networks impose specific design rules that we found necessary for either scheme to train at all. (i) Monte Carlo dropout~\citep{GalGhahramani2016}: dropout is left active at inference, $T$ stochastic forward passes are averaged, and their empirical variance is reported per pixel. Because perturbations are amplified by $\omega_0$ at every subsequent sine layer, aggressive dropout ($p=0.1$ after every layer) scrambles the phase structure and prevents learning; we use $p=0.05$ on the hidden activations only, with the first sine layer left deterministic. (ii) Mean-field variational inference~\citep{Blundell2015,Yang2021BPINN}: each weight is modeled as an independent Gaussian $\mathcal{N}(\mu_k,\sigma_k^2)$, trained by minimizing the evidence lower bound with the reparameterization trick, with $T$ posterior samples drawn at inference. Here the prior scale is critical; a generic prior ($\sigma_{\text{prior}}=0.1$) exceeds the SIREN weight scale $\sqrt{6/d_\ell}/\omega_0$ by more than an order of magnitude, and the KL term increases the weight noise until the network output becomes pure noise. We match the prior of each layer to its initialization scale and initialize the posterior noise well below it ($\sigma_k\approx10^{-4}$). Both variants are trained with the same normalized, ramped composite loss of Eq.~(\ref{eq:total}) and produce per-pixel mean and standard-deviation maps without retraining the deterministic model.

\begin{figure*}[t]
\centering
\includegraphics[width=\textwidth]{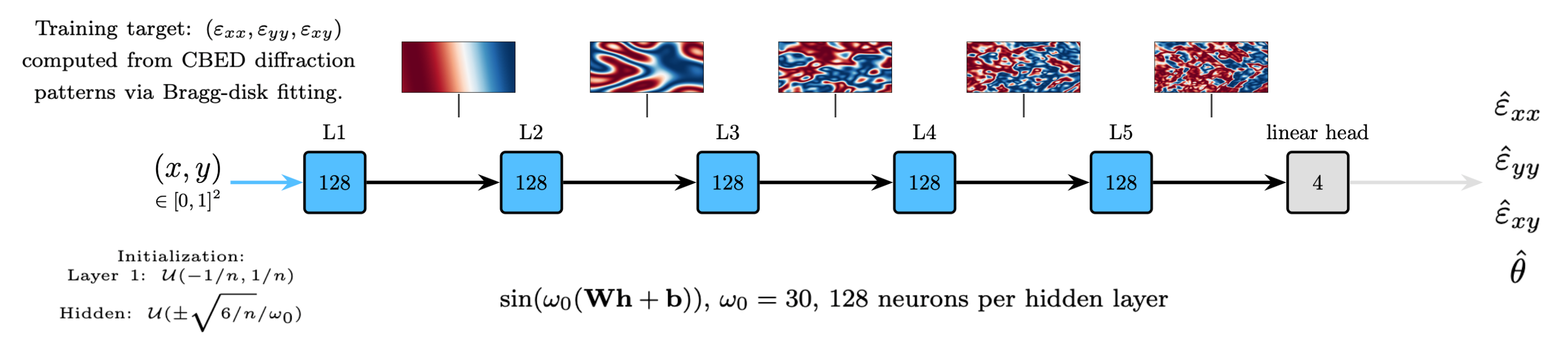}
\caption{SIREN backbone used for the reconstruction (Section~\ref{sec:siren}). A coordinate query $(x,y)\in[0,1]^2$ passes through sine layers of width $128$ (Eq.~(\ref{eq:siren}), $\omega_0=30$) and a final linear head returns the four field channels $(\hat\varepsilon_{xx},\hat\varepsilon_{yy},\hat\varepsilon_{xy},\hat\theta)$. The tiles above the layers illustrate the learned representation: first-layer neurons are plane-wave gratings $\sin(\omega_0(\mathbf{w}\cdot\mathbf{x})+b)$ whose orientation and frequency are set by the weights, and successive layers compose them into progressively structured interference patterns that converge on the chevron strain motif of the specimen (tiles are schematic renderings of this progression).}
\label{fig:arch}
\end{figure*}

%----------------------------------------------------------------------
\subsection{Evaluation metrics}
\label{sec:metrics}
%----------------------------------------------------------------------
For each component $c\in\{\varepsilon_{xx},\varepsilon_{yy},\varepsilon_{xy},\theta\}$ we report, computed against the full valid-pixel grid, the mean-squared error MSE, root MSE (RMSE), mean absolute error (MAE), peak absolute error, and the coefficient of determination $R^{2}=1-\text{SS}_{\text{res}}/\text{SS}_{\text{tot}}$. RMSE is also normalized by the dynamic range and the standard deviation of the measured field. Two physics-consistency metrics are unique to physics-informed reconstruction:
\begin{equation}
\mathcal{R}_{\text{equil}} \;=\; \sqrt{\mathcal{L}_{\text{equil}}(\hat{\boldsymbol{\phi}})},\qquad
\mathcal{R}_{\text{compat}} \;=\; \sqrt{\mathcal{L}_{\text{compat}}(\hat{\boldsymbol{\phi}})},
\end{equation}
each averaged over a dense evaluation grid. They quantify consistency with the imposed equilibrium and compatibility constraints independently of how well the reconstruction fits the possibly noisy data.

%======================================================================
\section{Implementation with a SIREN backbone and worked example}
\label{sec:impl}
%======================================================================

This section presents a specific realization of the architecture described in Section~\ref{sec:arch}. The chosen backbone is a sine-activated residual network (SIREN); the worked example is a single experimental 4D-STEM strain map. The map is used purely as a test bed for the algorithm; no conclusions are drawn about the material itself, and the same architecture and training protocol apply unchanged to 4D-STEM strain maps of semiconductor channels, heteroepitaxial films, 2D materials, oxides, etc.

%----------------------------------------------------------------------
\subsection{SIREN backbone}
\label{sec:siren}
%----------------------------------------------------------------------
Each hidden layer of the SIREN backbone implements
\begin{equation}
\mathbf{h}_{\ell+1} \;=\; \sin(\omega_0\,\mathbf{W}_\ell\mathbf{h}_\ell + \mathbf{b}_\ell),
\label{eq:siren}
\end{equation}
with frequency parameter $\omega_0$ governing spectral bandwidth. The choice of Eq.~(\ref{eq:siren}) is dictated by requirement 2 of Section~\ref{sec:requirements}: $\sin$ is $C^\infty$, so $f_{\boldsymbol{\phi}}$ is $C^\infty$; second-order spatial derivatives are evaluable to machine precision with no saturation. Three properties of Eq.~(\ref{eq:siren}) carry over directly to strain reconstruction.

First, the frequency parameter $\omega_0$ sets a preferred band of representable functions, and its choice is not free. With coordinates normalized to the unit square, $\omega_0=30$ (the value recommended by \citet{Sitzmann2020}) resolves the $\sim1$--$30$ pixel features of the strain map; we verified that $\omega_0=1$ renders the network too smooth to fit the data at any sampling fraction ($R^2\leq0.05$), while $\omega_0=60$ destabilizes optimization entirely. Second, the variance-preserving initialization of \citet{Sitzmann2020}, in which weights are drawn from $\mathcal{U}(-\sqrt{6/d_\ell}/\omega_0,\,+\sqrt{6/d_\ell}/\omega_0)$ for hidden layers of width $d_\ell$, keeps the activation distribution stationary across depth and avoids the gradient-collapse pathology of naive deep sine networks. Third, residual skip connections~\citep{He2016} every two layers further stabilize second-derivative backpropagation.

We use an input sine layer $\mathbb{R}^{2}\!\to\!\mathbb{R}^{128}$ followed by four hidden sine layers of width $d_\ell=128$, $\omega_0=30$, a linear skip projection entering the third hidden layer, and a final linear head $\mathbb{R}^{d_L}\to\mathbb{R}^4$ with no activation, allowing unrestricted sign on every component of $\mathbf{u}$. Figure~\ref{fig:arch} summarizes the backbone and the layer-by-layer construction of the field. The total parameter count is $83{,}460$. These hyperparameters were fixed once and held constant across all experiments below. (Wider variants are learning-rate fragile: $d_\ell=256$ diverges at the default $10^{-3}$ and requires $\sim5\times10^{-4}$, without improving accuracy.)

%----------------------------------------------------------------------
\subsection{Example specimen and dataset}
\label{sec:specimen}
%----------------------------------------------------------------------
The test dataset is a 4D-STEM strain map of the IV--VI high-entropy thermoelectric compound PbGeSnSe$_{1.5}$Te$_{1.5}$, from the domain-structure study of \citet{liu2024}, exhibiting extended chevron-shaped strain bands associated with its ferroelastic domain structure. The 4D-STEM data were recorded on an aberration-corrected STEM with a pixelated direct electron detector, and Bragg-disk positions were extracted by a standard cross-correlation template and converted to the 2D strain tensor in the standard way~\citep{Padgett2020,Savitzky2021}. The dataset is a $180\times400 = 72{,}000$ pixel grid ($17.9$\,nm/pixel) with all pixels passing Bragg-disk validity; the measured channels are $(\varepsilon_{xx},\varepsilon_{yy},\varepsilon_{xy})$, and the fourth channel $\theta=\tfrac{1}{2}\operatorname{atan2}\!\left(2\varepsilon_{xy},\varepsilon_{xx}-\varepsilon_{yy}\right)$ is the derived principal-strain orientation rather than an independent measurement; its reconstruction scores should be read accordingly. 

\begin{figure}[t]
\centering
\includegraphics[width=0.95\linewidth]{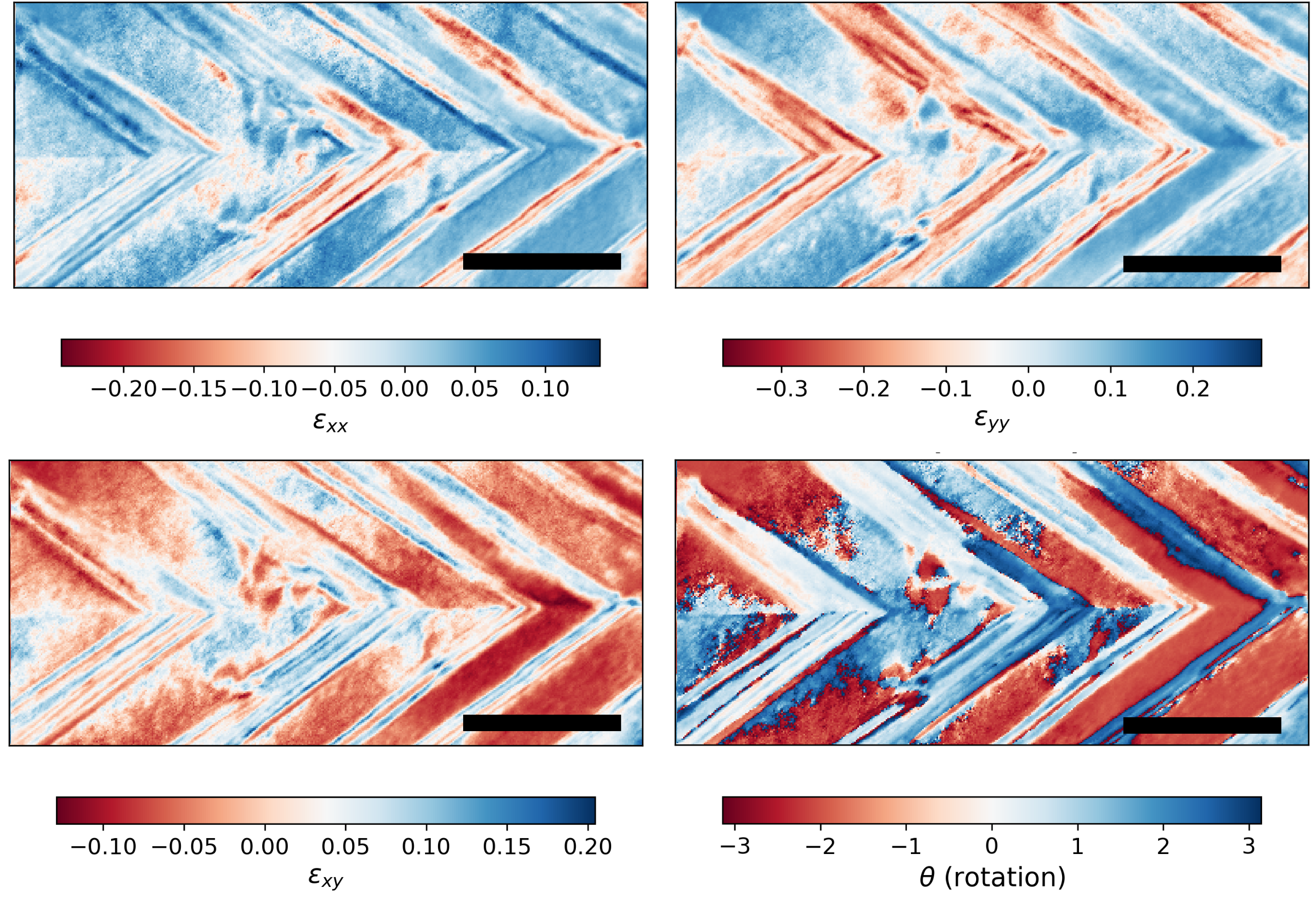}
\caption{Experimental 4D-STEM strain map ($180\times400$ pixels, all valid) used as the worked example. Panels show the three measured strain channels ($\varepsilon_{xx}$, $\varepsilon_{yy}$, $\varepsilon_{xy}$), the derived angle $\theta$, the valid-pixel mask, and the per-pixel strain-magnitude map. Chevron-shaped strain bands are superposed on pixel-scale measurement noise ($\sim3\%$ of the field variance). Scale bar is 300\,nm.}
\label{fig:raw}
\end{figure}

The empirical ranges are $\varepsilon_{xx}\in[-0.24,\,0.14]$, $\varepsilon_{yy}\in[-0.37,\,0.28]$ and $\varepsilon_{xy}\in[-0.13,\,0.20]$. A $3\times3$ median-filter decomposition attributes $\sim97\%$ of the field variance to spatially extended structure and $\sim3\%$ to pixel-scale measurement noise, which bounds the achievable $R^2$ against the raw reference at $\approx0.97$. The physics terms use isotropic plane-stress elasticity with representative constants $E=150$\,GPa and $\nu=0.27$. Under the scale normalization of Eq.~(\ref{eq:total}), the common multiplicative modulus factor cancels from the training objective; consequently, the effective Poisson-ratio dependence, rather than the absolute value of $E$, governs the equilibrium prior. Plane stress is adopted as an effective thin-foil approximation. For specimens whose geometry or material symmetry warrants it, the equilibrium residual can instead use plane strain, anisotropic elasticity, spatially heterogeneous elastic parameters, or an eigenstrain-augmented constitutive closure.

Figure~\ref{fig:raw} presents the four measurement channels together with the corresponding per-pixel strain-magnitude map. This dataset serves as a stringent benchmark for evaluating the robustness and generalization capacity of the proposed architecture. The extended bands probe its ability to achieve smooth, coherent reconstructions across approximately two orders of magnitude variation in sampling density, whereas the pixel-scale noise challenges the imposed priors in their role as effective denoising mechanisms.

%----------------------------------------------------------------------
\subsection{Sparse sampling protocol}
\label{sec:sampling}
%----------------------------------------------------------------------
\begin{figure}[t]
\centering
\includegraphics[width=\linewidth]{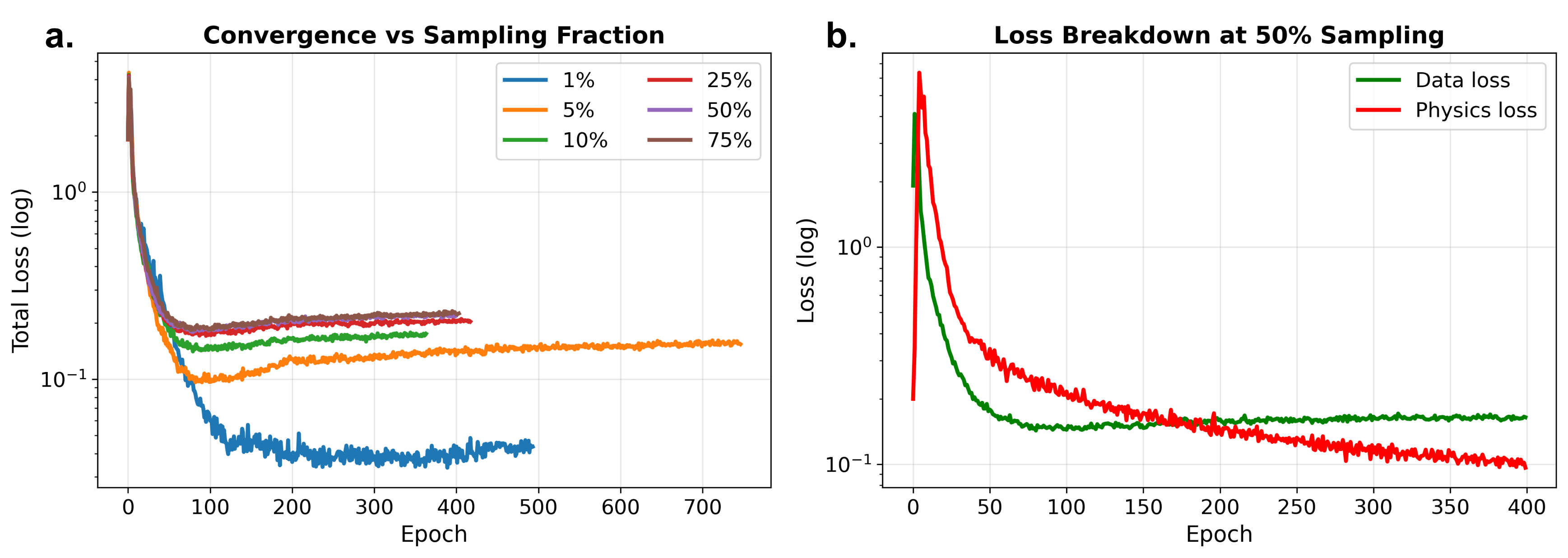}\\[4pt]
\caption{(a) Representative random training-point distributions for the six sampling fractions. (b) Total-loss trajectories across sampling fractions. Training was terminated by validation-loss early stopping after 364--747 epochs, and all reported reconstructions and metrics use the restored best-validation checkpoint. (c) Decomposition of the total loss at 50\% sampling into data and scale-normalized physics components.}
\label{fig:sampling}
\end{figure}

To emulate dose-reduced acquisition we subsampled the full map uniformly at random to retain a fraction $f\in\{1\%,5\%,10\%,25\%,50\%,75\%\}$ of the valid pixels ($720$, $3{,}600$, $7{,}200$, $18{,}000$, $36{,}000$ and $54{,}000$ probe positions). Figure~\ref{fig:sampling}(a) shows representative training-point distributions; Fig.~\ref{fig:sampling}(b) shows loss trajectories and the data/physics breakdown at $50\%$ sampling.

%----------------------------------------------------------------------
\subsection{Reconstructed strain tensor}
%----------------------------------------------------------------------
\begin{figure*}[t]
\centering
\includegraphics[width=\textwidth]{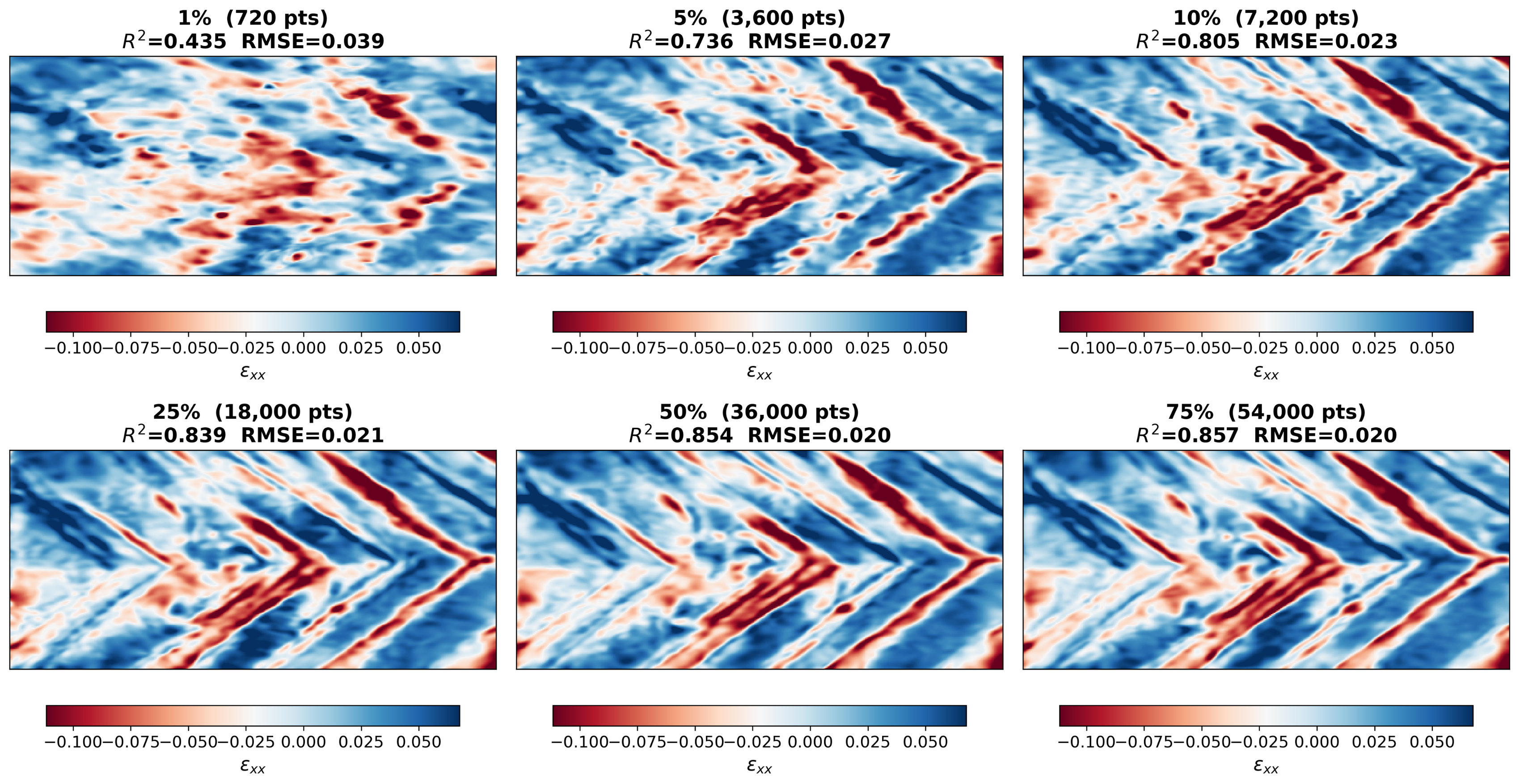}
\caption{PINN reconstructions of $\varepsilon_{xx}$ at the six sampling fractions, with retained training points overlaid in green. The chevron strain bands are recovered from $10\%$ of the probe positions; fine structure is progressively restored as sampling increases and saturates by $\sim25\%$. Panel titles report $R^2$ and RMSE against the full valid-pixel grid. Scale bar is 300\,nm.}
\label{fig:pred_exx}
\end{figure*}

Figure~\ref{fig:pred_exx} shows the reconstructed $\varepsilon_{xx}$ field at each sampling fraction, with retained training points overlaid in green. Two qualitative features stand out. First, the chevron strain-band morphology is recovered from remarkably sparse data: at $10\%$ sampling ($7{,}200$ probes, corresponding to an approximately tenfold reduction in nominal probe-dose budget at fixed current and dwell time) the bands are reproduced with the correct sign, magnitude and geometry, and even $720$ probes ($1\%$) recover the large-scale strain distribution, though not the individual bands. Second, fine structure is progressively restored as sampling increases, saturating by $\sim25\%$: beyond that point the residual disagreement with the raw reference is dominated by pixel-scale measurement noise that a reconstruction intended to recover the underlying smooth strain field need not reproduce. The $\varepsilon_{yy}$ and $\varepsilon_{xy}$ channels behave equivalently (Supplementary Figs.~S1--S2).

%----------------------------------------------------------------------
\subsection{Quantitative metrics}
\begin{figure}[t]
\centering
\includegraphics[width=\linewidth]{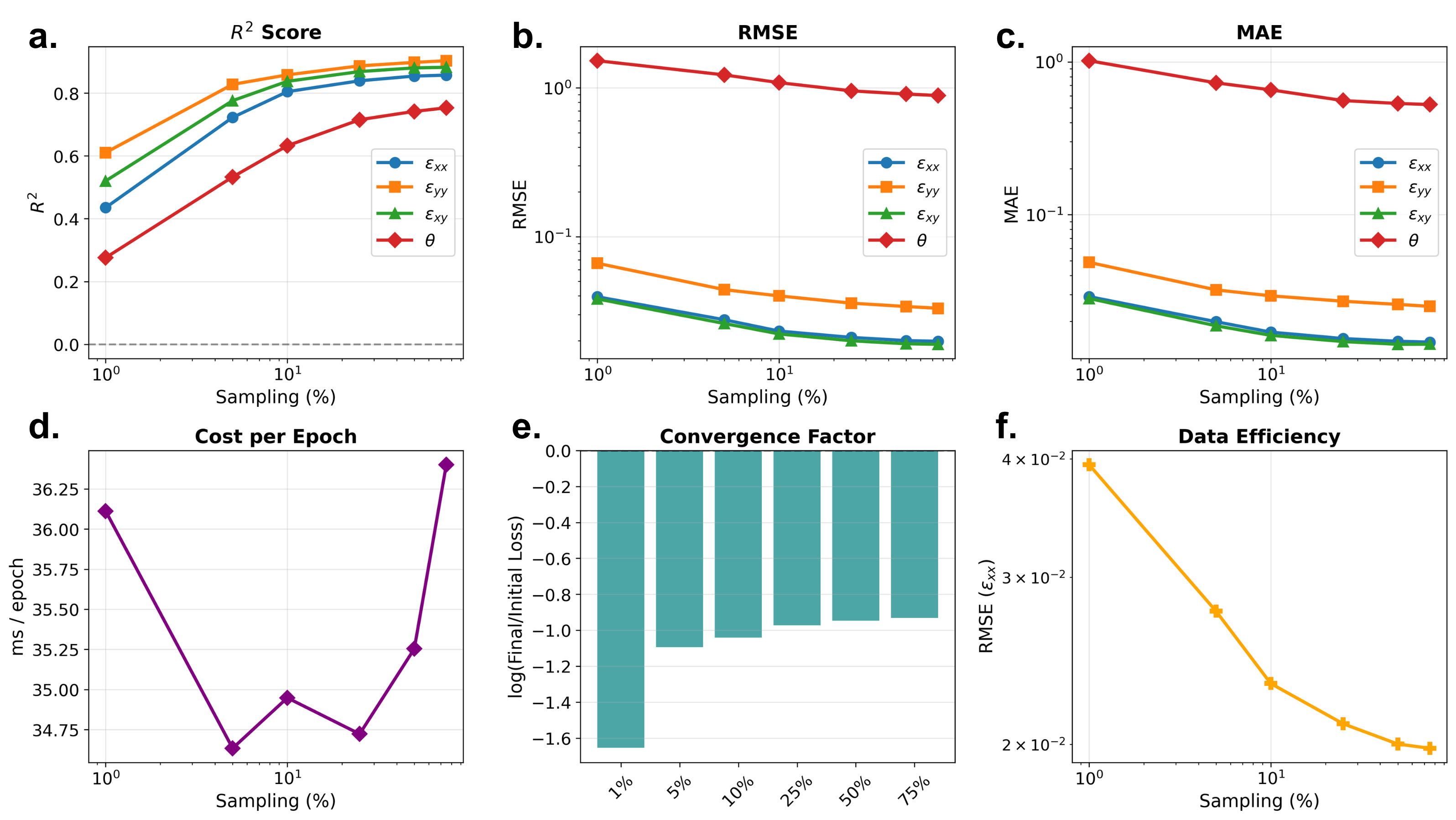}
\caption{Reconstruction metrics versus sampling fraction. a. $R^2$, b. RMSE, c. MAE per strain component, d. training wall-clock time, e. loss-reduction convergence factor, and f. data-efficiency curve showing $\varepsilon_{xx}$ RMSE versus sampling fraction on a log--log axis.}
\label{fig:metrics}
\end{figure}

%----------------------------------------------------------------------
Table~\ref{tab:metrics} and Fig.~\ref{fig:metrics} report reconstruction metrics against the full valid-pixel grid. RMSE on $\varepsilon_{xx}$ lies in $(2.0$--$3.9)\times 10^{-2}$ across sampling fractions; MAE is in the tighter $(1.5$--$2.9)\times 10^{-2}$ band and decreases to $1.46\times 10^{-2}$ at $75\%$ sampling. The $R^2$ on $\varepsilon_{xx}$ rises steeply from $0.44$ at $1\%$ to $0.80$ at $10\%$ and then saturates ($0.84$ at $25\%$, $0.86$ at $75\%$); $\varepsilon_{yy}$ and $\varepsilon_{xy}$ reach $R^2 = 0.90$ and $0.88$ at $75\%$. The saturation is expected: pixel-scale measurement noise accounts for $\sim3\%$ of the reference variance and the sharpest band edges are unresolvable at sparse probe spacing, so $10$--$25\%$ of the probe positions capture essentially all recoverable structure.

\begin{figure}[t]
\centering
\includegraphics[width=\linewidth]{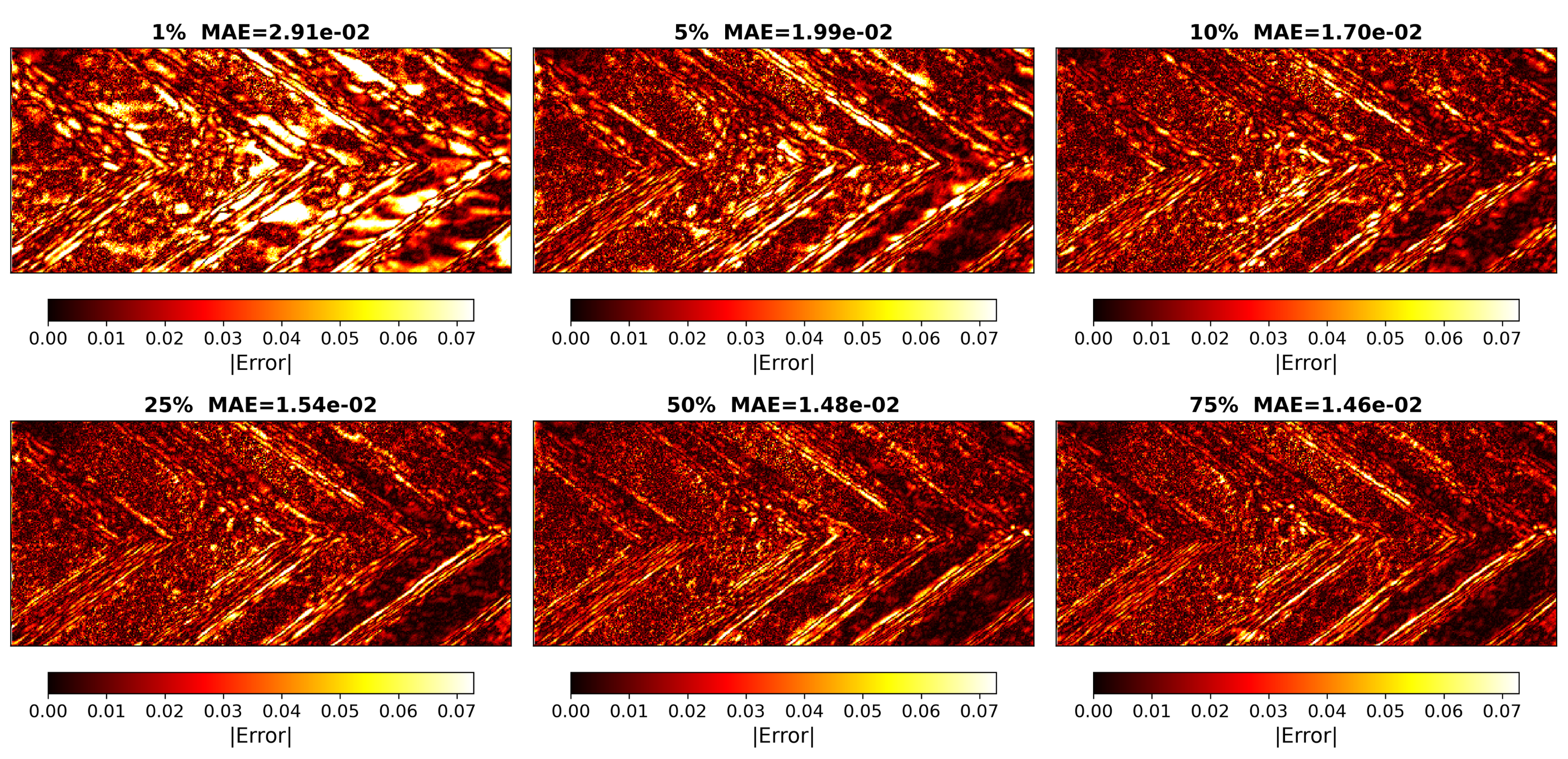}
\caption{Absolute-error heatmaps for $\varepsilon_{xx}$ at the six sampling fractions. Error decreases rapidly up to $\sim10\%$ sampling and is thereafter concentrated on pixel-scale noise and the sharpest band edges, which the physics-regularized network does not attempt to reproduce.}
\label{fig:errors}
\end{figure}

Even at $1\%$ sampling ($720$ probe positions), all four components maintain $R^2 > 0.27$ and the large-scale strain distribution is resolved in the mean map. The spatial error maps in Fig.~\ref{fig:errors} confirm that the dominant error contribution at moderate-to-dense sampling is pixel-scale noise together with the sharpest band edges.

\begin{table*}[t]
\caption{Reconstruction metrics for the worked-example specimen versus sampling fraction, computed against the full valid-pixel grid (lower is better for RMSE and MAE; higher is better for $R^2$).}
\label{tab:metrics}
\centering
\begin{tabular*}{\textwidth}{@{\extracolsep{\fill}}lcccccccc@{}}
\toprule
$f$ & \# pts & $R^2(\varepsilon_{xx})$ & RMSE($\varepsilon_{xx}$) & MAE($\varepsilon_{xx}$) & $R^2(\varepsilon_{yy})$ & $R^2(\varepsilon_{xy})$ & $R^2(\theta)$ & $R^2_\text{avg}$ \\
\midrule
 1\% &     720 & $+0.44$ & $3.9\times10^{-2}$ & $2.9\times10^{-2}$ & $+0.61$ & $+0.52$ & $+0.28$ & $+0.46$ \\
 5\% &   3,600 & $+0.72$ & $2.8\times10^{-2}$ & $2.0\times10^{-2}$ & $+0.83$ & $+0.78$ & $+0.53$ & $+0.71$ \\
10\% &   7,200 & $+0.80$ & $2.3\times10^{-2}$ & $1.7\times10^{-2}$ & $+0.86$ & $+0.84$ & $+0.63$ & $+0.78$ \\
25\% &  18,000 & $+0.84$ & $2.1\times10^{-2}$ & $1.5\times10^{-2}$ & $+0.89$ & $+0.87$ & $+0.72$ & $+0.83$ \\
50\% &  36,000 & $+0.85$ & $2.0\times10^{-2}$ & $1.5\times10^{-2}$ & $+0.90$ & $+0.88$ & $+0.74$ & $+0.84$ \\
75\% &  54,000 & $+0.86$ & $2.0\times10^{-2}$ & $1.5\times10^{-2}$ & $+0.90$ & $+0.88$ & $+0.75$ & $+0.85$ \\
\bottomrule
\end{tabular*}
\end{table*}

%----------------------------------------------------------------------
\subsection{Physics-consistency diagnostics}
%----------------------------------------------------------------------
Consistency with the imposed mechanical constraints is scored in physical units by evaluating $\mathcal{R}_\text{equil}$ and $\mathcal{R}_\text{compat}$ with central finite differences on the reconstructed $17.9$\,nm/pixel grid, so that networks and non-network baselines are scored by the same operator. The raw measured map itself carries $\mathcal{R}_\text{equil}=300$\,GPa\,$\mu$m$^{-1}$ and $\mathcal{R}_\text{compat}=54\,\mu$m$^{-2}$, dominated by pixel-scale noise. PINN reconstructions reduce these to $110$--$164$\,GPa\,$\mu$m$^{-1}$ and $9.7$--$15.4\,\mu$m$^{-2}$ across sampling fractions. An ablation in which the same network is trained with $\lambda_{\text{phys}}=0$ (a data-only SIREN baseline of identical architecture and schedule) produces residuals that are $1.5$--$1.8\times$ (equilibrium) and $2.2$--$3.9\times$ (compatibility) larger than the PINN's at every sampling fraction. At equal data, equal capacity and equal training cost, the physics term purchases a consistent improvement in self-consistency at no inference-time penalty; its effect on reconstruction \emph{accuracy} is regime-dependent and is quantified in the Discussion.

%----------------------------------------------------------------------
\subsection{Bayesian epistemic uncertainty}
\begin{figure*}[t]
\centering
\includegraphics[width=\textwidth]{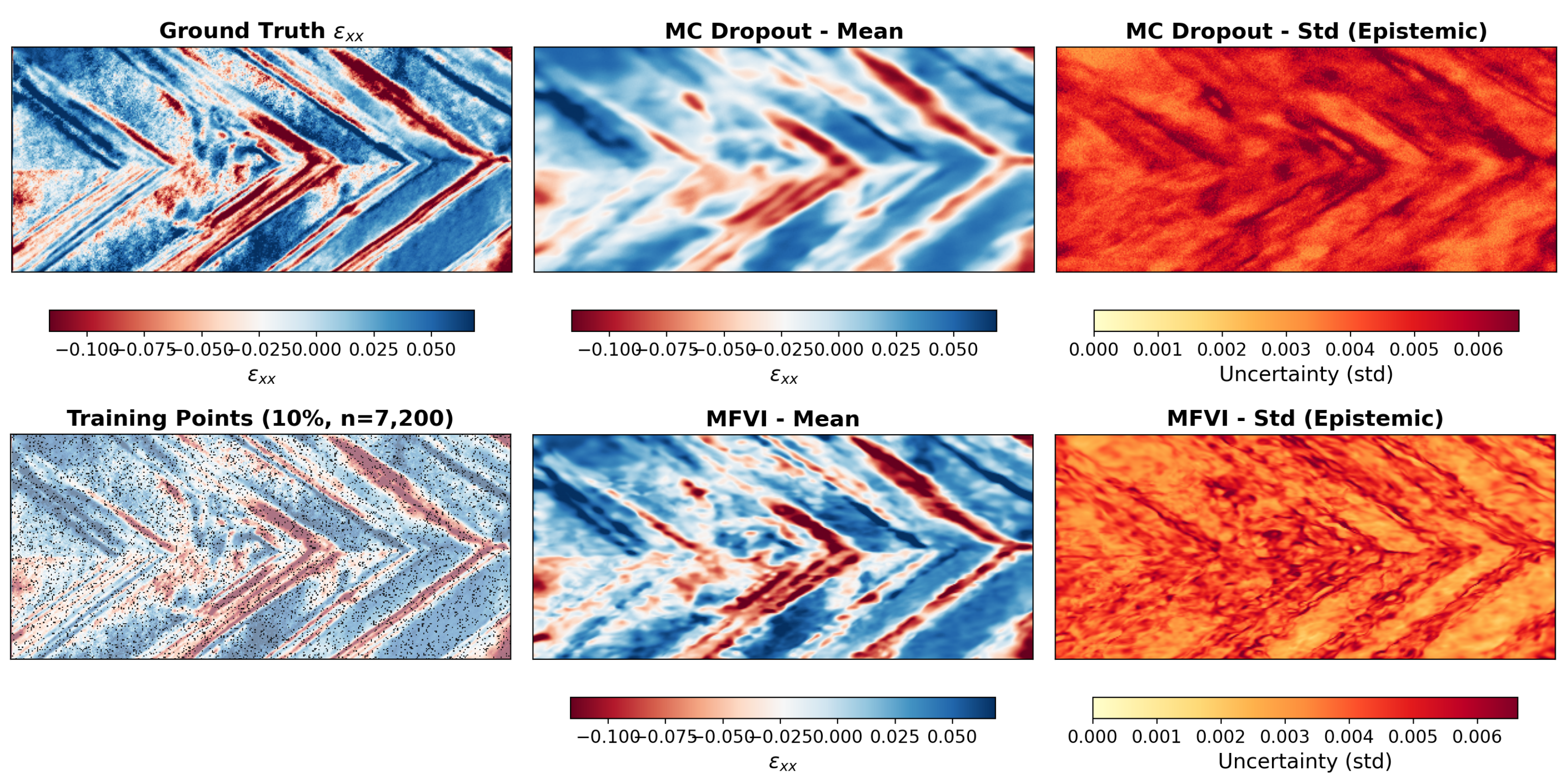}
\caption{Bayesian epistemic-uncertainty maps for $\varepsilon_{xx}$ at $10\%$ sparse data, $T=150$ stochastic forward passes. Top row: ground-truth field; Monte Carlo dropout mean and standard deviation. Bottom row: training-point distribution; mean-field variational-inference mean and standard deviation. Both posterior means recover the chevron strain bands; the uncertainty maps (in strain units) are spatially structured and concentrate on the bands and band edges, where reconstruction error is largest.}
\label{fig:uq}
\end{figure*}

%----------------------------------------------------------------------
Figure~\ref{fig:uq} shows per-pixel epistemic uncertainty from MCD and MFVI at $10\%$ sampling ($T=150$ stochastic passes; $7{,}200$ training pixels). Both variants recover the chevron strain-band morphology in their posterior means, and both produce spatially structured uncertainty maps ($\sigma\approx3$--$6\times10^{-3}$ strain, reported in the same units as the mean) that concentrate on the strain bands and band edges, i.e., the high-gradient regions where posterior uncertainty is expected to be elevated. In held-out validation the MFVI uncertainty correlates with the actual per-pixel reconstruction error ($\rho\approx0.32$; MCD $\rho\approx0.27$), and the MFVI mean is the most accurate single reconstruction in this study ($R^2(\varepsilon_{xx})=0.82$ at $10\%$ sampling). Operationally, the uncertainty map is a target for adaptive acquisition: a follow-up sparse scan can concentrate probes on the high-uncertainty regions identified after a first pass.

%----------------------------------------------------------------------
\subsection{Wall-clock cost}
%----------------------------------------------------------------------
Training terminated by early stopping after $364$--$747$ epochs of the $5{,}000$-epoch budget, with the best-validation weights restored. The computational cost per epoch is dominated by the second-order autodiff of the physics terms and is nearly constant at $\approx35$\,ms on Apple Silicon M-series MPS regardless of sampling fraction, giving total wall-clock times of $13$--$26$\,s per model; the entire six-fraction study, ablations, classical baselines and both Bayesian variants complete in $\sim10$ minutes on a laptop. Inference over the full $72{,}000$-pixel grid costs $\lesssim 1$\,s. Peak GPU memory is below $2$\,GB per model.

%======================================================================
\section{Discussion}
%======================================================================

\subsection{Interpretation of the PINN reconstruction}
The network simultaneously fulfills two distinct functions. On the strain bands, it acts as a reconstruction algorithm. It reconstructs a smooth strain tensor that better satisfies the prescribed equilibrium and compatibility constraints from a strongly undersampled set of measurements, and its reconstruction error decreases in a controlled way as the sampling fraction increases. On the pixel-scale noise it acts as a denoiser. The smooth implicit \emph{ansatz} combined with the physics terms damps the high-spatial-frequency components of the raw data, yielding a field that is likely closer to the true lattice strain than the noisy reference. Scoring against the raw reference therefore understates the reconstruction quality: pixel-scale noise alone caps the attainable $R^2$ at $\approx0.97$, and the observed saturation of $R^2$ near $0.86$ beyond $25\%$ sampling reflects this ceiling together with the sharpest band edges, not a failure of convergence (training longer, widening the network, or raising $\omega_0$ does not improve it).

\subsection{Comparison with alternative sparse reconstructions, and when the physics prior helps}
Compressed sensing with an $\ell_1$-on-wavelet prior (ISTA), Gaussian-process regression with a Mat\'ern-$3/2$ kernel, and a SIREN baseline with $\lambda_{\text{phys}}=0$ were all evaluated on the same sampling masks. At $10\%$ sampling the PINN reduces MAE on $\varepsilon_{xx}$ by $26\%$ relative to CS and $22\%$ relative to GP, while carrying equilibrium residuals $1.4\times$ lower than CS.

The ablation relative to the data-only SIREN reveals a more complex picture. The physics prior improves accuracy only in the extreme-sparsity regime: at $1\%$ sampling it lowers MAE by $\approx10\%$, at $5\%$ it is neutral, and from $10\%$ upwards the data-only SIREN is more accurate (by $10\%$ at $10\%$ sampling, growing to $86\%$ at $75\%$), even though the PINN always produces the more physically self-consistent field. This is the expected behavior for a ferroelastic, domain-structured crystal. The measured lattice strain contains the spontaneous (transformation) eigenstrain associated with the domains~\citep{liu2024}. As a result, homogeneous isotropic equilibrium, Eq.~(\ref{eq:equilibrium}) with constant $(\lambda,\mu)$, is only an approximate prior on it, an excellent regularizer when data are too sparse to constrain the field, and an increasing source of bias once they are not. The practical protocol that follows is simple: enable the physics terms in proportion to sparsity, and treat a growing PINN-vs-data-only gap as evidence of unmodeled physics (here, domain eigenstrain) rather than as a defect of either method. None of the non-PINN methods produce reconstructions that satisfy compatibility to within the same tolerance, which matters whenever downstream analysis requires derivative quantities such as rotation, shear gradient or defect-free displacement.

\subsection{Practical implications for 4D-STEM strain experiments}
The actionable consequence is a dose-reduction protocol that is agnostic to material class. Under fixed probe current and dwell time, random retention of 10\% of probe positions reduces the nominal probe-dose budget by approximately tenfold while preserving reconstruction of the physically informative features (here, the chevron strain bands) to within $2.3\times 10^{-2}$ RMSE and $1.7\times 10^{-2}$ MAE in $\varepsilon_{xx}$ ($R^2 = 0.80$). With a constitutive model appropriate to the specimen, the same framework can be adapted to strain characterization in semiconductor channels, heteroepitaxial films, 2D materials, ferroic oxides, and functional battery and thermoelectric phases. The Bayesian uncertainty map provides a target for adaptive acquisition: a second-pass scan can concentrate probes on the high-uncertainty regions identified after a first sparse pass, converging on the acquisition plan that is optimal for the specimen on the holder.

\subsection{Limitations}
Six caveats. First, the physics terms assume isotropic Hooke's law with constant elastic constants; the ablation shows this is only approximate for the present domain-structured specimen, and strongly anisotropic, compositionally graded, or eigenstrained specimens require per-pixel, learnable, or eigenstrain-augmented constitutive closures~\citep{Tartakovsky2020,Henkes2022}. Second, the physics residuals are formed on per-component standardized fields; unequal component variances mildly distort the effective Poisson ratio in the operator, and a physical-unit formulation would be cleaner. Third, per-component standardization statistics were estimated from the dense reference map for this retrospective benchmark, constituting a mild information leak relative to a strictly prospective sparse acquisition. A prospective implementation should estimate these statistics from retained measurements only. Fourth, $\theta$ is derived from the strain components rather than measured independently. Fifth, the GP baseline is capped at $1{,}500$ training points for tractability, so GP comparisons at $\geq25\%$ sampling understate its performance. Sixth, discontinuous strain fields (cracks, grain boundaries, dislocation cores) violate the differentiability assumptions baked into SIREN; domain-decomposition PINNs and discontinuity-adapted architectures are active areas of research. The reported study uses one dataset and one random mask per sampling fraction; repeated mask realizations would attach error bars to Table~\ref{tab:metrics}.

\subsection{Extensions}

The framework is portable to other 4D-STEM modalities by replacing the mechanical PDE residuals with modality-specific forward models and constraints. In ptychography, introduced by Hoppe as a phase-retrieval scheme~\citep{Hoppe1969a,Hoppe1969b,HoppeStrube1969}, the measurement model couples a scanned probe to a complex transmission function under coherent paraxial propagation; thin-object reconstructions use a multiplicative interaction model, while thick or strongly scattering specimens require multislice variants with alternating transmission and Fresnel propagation~\citep{RodenburgMaiden2019}. In differential phase contrast (DPC), the signal provides a projected in-plane field (or phase gradient) that can be integrated to obtain a scalar potential under appropriate electrostatic and projection assumptions, and whose divergence can be related to projected charge density through Gauss's law~\citep{Lubk2015,shibata2015imaging}. Three-dimensional tomographic strain reconstruction similarly inherits equilibrium and compatibility in volumetric form. Across these settings, physics-informed objectives can reduce dependence on large labeled training sets by embedding the relevant forward physics and constitutive constraints directly into the reconstruction loss.
%======================================================================
\section{Conclusion}
%======================================================================
We have developed a PINN architecture for 4D-STEM strain reconstruction in which the architectural choices are driven by the priors. Five ingredients are required: a coordinate-based implicit \emph{ansatz}, an activation with non-vanishing second derivatives, scale normalization of the physics residuals frozen after a short warmup, an exponential physics-weight ramp, and residual-based adaptive collocation. The architecture is mesh-free and differentiable, and can incorporate constitutive closures appropriate to different material systems while returning the in-plane strain tensor and derived orientation field at arbitrary real-space coordinates. Instantiated with a SIREN backbone and trained on an experimental $180\times400$-pixel strain map of the IV--VI high-entropy thermoelectric PbGeSnSe$_{1.5}$Te$_{1.5}$, the method recovers the chevron strain-band morphology across sampling fractions spanning two orders of magnitude ($1$--$75\%$, corresponding to $720$--$54{,}000$ probe positions). MAE on $\varepsilon_{xx}$ stays in $(1.5$--$2.9)\times 10^{-2}$ across the full sampling range; $R^2$ rises from $0.44$ at $1\%$ to $0.80$ at $10\%$ and saturates near $0.86$, at the ceiling set by pixel-scale measurement noise. At $10\%$ sampling the method outperforms compressed sensing and Gaussian-process baselines by $26\%$ and $22\%$ in MAE. The ablation against a data-only SIREN of identical capacity delineates when the physics prior pays: it improves accuracy in the extreme-sparsity regime and always improves physical self-consistency (equilibrium residuals $1.5$--$1.8\times$ lower, compatibility $2.2$--$3.9\times$ lower), but biases the reconstruction once data suffice --- consistent with domain eigenstrain rendering homogeneous equilibrium approximate for ferroelastic microstructures. Bayesian variants furnish per-pixel epistemic-uncertainty maps whose magnitude tracks the actual error. Together, these properties support approximately tenfold reductions in nominal probe-dose budget under fixed-current, fixed-dwell-time acquisition, with a spatially resolved uncertainty budget attached to every reconstructed pixel.

%======================================================================
\section{Competing interests}
No competing interest is declared.

\section{Author contributions statement}

R.dR. conceived the study, developed the overall methodology, directed the analysis, and drafted the manuscript. R.dR. and G.T.dS. designed and implemented the PINN architectures, training pipelines, and computational analyses. Y.L. contributed to the experimental design and interpretation of the 4D-STEM strain-mapping data. X.H. and V.P.D. contributed to scientific discussions and interpretation of the microscopy and materials science context. All authors contributed to manuscript revision and approved the final version.

\section{Data availability}
The processed strain-tensor maps used as the worked example, the full training and evaluation pipeline (a single Jupyter notebook, with a Colab-ready GPU twin), the configuration, and the scripts that regenerate every figure and table in this paper are available at \url{https://github.com/rmsreis/pinns-4dstem}. Reported metrics are written by the notebook to \nolinkurl{outputs/sota_adaptive-2/} as comma-separated files (\nolinkurl{results_summary.csv}, \nolinkurl{detailed_metrics.csv}, \nolinkurl{physics_diagnostics.csv}, \nolinkurl{ablation_summary.csv}, \nolinkurl{baselines_summary.csv}). The raw 4D-STEM diffraction dataset is available from the corresponding author on reasonable request.

\section{Acknowledgments}
This work made use of the EPIC facility ($RRID: SCR\textunderscore 026361$) of Northwestern University’s NU\textit{ANCE} Center, which has received support from the IIN and Northwestern's MRSEC program (NSF DMR-2308691).This research was supported in part through the computational resources by the Quest high performance computing facility at Northwestern University, jointly supported by the Office of the Provost, Office for Research, and Northwestern University Information Technology.

\bibliographystyle{unsrtnat}
\bibliography{reference}

%======================================================================
\clearpage
\appendix
\onecolumn
\section*{Supplementary Material}
\addcontentsline{toc}{section}{Supplementary Material}
%======================================================================

% Reset float numbering in the Supplementary Material
\setcounter{table}{0}
\renewcommand{\thetable}{S\arabic{table}}
\setcounter{figure}{0}
\renewcommand{\thefigure}{S\arabic{figure}}

This Supplementary Material reports the quantitative tables that support the baseline, ablation, and physics-consistency claims in the main text. All values were assembled from the notebook outputs \nolinkurl{results_summary-5.csv}, \nolinkurl{baselines_summary-2.csv}, and \nolinkurl{ablation_summary.csv} (physics-consistency diagnostics for the PINN are the $\mathcal{R}^{\mathrm{PINN}}_{\mathrm{equil}}$/$\mathcal{R}^{\mathrm{PINN}}_{\mathrm{compat}}$ columns of Table S3, identical to \nolinkurl{physics_diagnostics.csv}).

\subsection*{Supplementary Table S1. PINN summary versus sampling fraction}

% Tighter float spacing for the Supplementary tables
\setlength{\textfloatsep}{8pt plus 2pt minus 2pt}
\setlength{\floatsep}{6pt plus 2pt minus 2pt}
\setlength{\intextsep}{8pt plus 2pt minus 2pt}
\captionsetup[table]{skip=4pt}

\begin{table}[H]
\centering
\caption{Supplementary Table S1. Summary of the PINN reconstruction versus sampling fraction. Sampling is the retained fraction of the full $180\times400$ valid-pixel grid; train points is the corresponding raw probe count; time is the end-to-end training time to early stopping.}
\begin{tabular}{lcccccc}
\toprule
Sampling & Train points & $R^2(\varepsilon_{xx})$ & RMSE$(\varepsilon_{xx})$ & $R^2_{\mathrm{avg}}$ & Time (s) & Epochs \\
\midrule
1.0\%  & 720   & 0.4353 & 3.9426e-02 & 0.4605 & 17.8 & 494 \\
5.0\%  & 3600  & 0.7227 & 2.7629e-02 & 0.7149 & 25.9 & 747 \\
10.0\% & 7200  & 0.8047 & 2.3185e-02 & 0.7833 & 12.7 & 364 \\
25.0\% & 18000 & 0.8393 & 2.1030e-02 & 0.8275 & 14.5 & 418 \\
50.0\% & 36000 & 0.8544 & 2.0020e-02 & 0.8436 & 14.1 & 400 \\
75.0\% & 54000 & 0.8574 & 1.9815e-02 & 0.8490 & 14.7 & 404 \\
\bottomrule
\end{tabular}
\end{table}

\subsection*{Supplementary Tables S2 and S3. Baseline comparison}

\begin{table}[H]
\centering
\caption{Supplementary Table S2. Baseline comparison for $\varepsilon_{xx}$ at 1\%, 5\%, and 10\% sampling. CS: compressed sensing; GP: Gaussian-process regression; SIREN-only: data-only network with $\lambda_{\mathrm{phys}}=0$; PINN: full physics-informed model.}
\begin{tabular}{llccccc}
\toprule
Sampling & Method & RMSE & MAE & $R^2$ & $\mathcal{R}_{\mathrm{equil}}$ & $\mathcal{R}_{\mathrm{compat}}$ \\
\midrule
1\%  & CS         & 0.0531014 & 0.0410219 & -0.0244 & 136.78 & 48.66 \\
1\%  & GP         & 0.0357366 & 0.0262034 &  0.5361 &  88.67 &  5.23 \\
1\%  & SIREN-only & 0.0437027 & 0.0322720 &  0.3062 & 201.77 & 33.60 \\
1\%  & PINN       & 0.0394263 & 0.0290726 &  0.4353 & 110.07 & 15.22 \\
\midrule
5\%  & CS         & 0.0456676 & 0.0321751 & 0.2424 & 232.81 & 81.05 \\
5\%  & GP         & 0.0313421 & 0.0218427 & 0.6431 & 133.99 &  9.95 \\
5\%  & SIREN-only & 0.0270346 & 0.0188362 & 0.7345 & 223.07 & 37.75 \\
5\%  & PINN       & 0.0276294 & 0.0199034 & 0.7227 & 124.59 &  9.74 \\
\midrule
10\% & CS         & 0.0349020 & 0.0229680 & 0.5575 & 222.77 & 71.74 \\
10\% & GP         & 0.0304912 & 0.0217676 & 0.6623 & 126.95 &  9.64 \\
10\% & SIREN-only & 0.0219998 & 0.0154348 & 0.8242 & 245.15 & 41.98 \\
10\% & PINN       & 0.0231854 & 0.0170203 & 0.8047 & 163.59 & 15.40 \\
\bottomrule
\end{tabular}
\end{table}

\begin{table}[H]
\centering
\caption{Supplementary Table S3. Baseline comparison for $\varepsilon_{xx}$ at 25\%, 50\%, and 75\% sampling. Residual diagnostics are reported in GPa/\textmu m and \textmu m$^{-2}$.}
\begin{tabular}{llccccc}
\toprule
Sampling & Method & RMSE & MAE & $R^2$ & $\mathcal{R}_{\mathrm{equil}}$ & $\mathcal{R}_{\mathrm{compat}}$ \\
\midrule
25\% & CS         & 0.0238530 & 0.0143634 & 0.7933 & 229.77 & 64.53 \\
25\% & GP         & 0.0310062 & 0.0217989 & 0.6508 & 126.99 &  8.21 \\
25\% & SIREN-only & 0.0153163 & 0.0111011 & 0.9148 & 259.91 & 42.71 \\
25\% & PINN       & 0.0210302 & 0.0154492 & 0.8393 & 153.76 & 12.45 \\
\midrule
50\% & CS         & 0.0146264 & 0.0075220 & 0.9223 & 252.32 & 59.47 \\
50\% & GP         & 0.0298390 & 0.0215583 & 0.6766 & 127.12 &  7.91 \\
50\% & SIREN-only & 0.0119739 & 0.0088095 & 0.9479 & 262.02 & 40.56 \\
50\% & PINN       & 0.0200196 & 0.0148280 & 0.8544 & 152.04 & 12.55 \\
\midrule
75\% & CS         & 0.0092112 & 0.0034514 & 0.9692 & 278.44 & 57.03 \\
75\% & GP         & 0.0299163 & 0.0212511 & 0.6749 & 109.04 & 10.43 \\
75\% & SIREN-only & 0.0104655 & 0.0078774 & 0.9602 & 264.91 & 38.55 \\
75\% & PINN       & 0.0198152 & 0.0146410 & 0.8574 & 155.32 & 12.05 \\
\bottomrule
\end{tabular}
\end{table}

\subsection*{Supplementary Table S4. PINN versus data-only SIREN ablation}

\begin{table}[H]
\centering
\caption{Supplementary Table S4. Ablation of the physics prior against a data-only SIREN. Ratios greater than one indicate lower residuals for PINN. MAE reduction is defined as $(\mathrm{MAE}_{\mathrm{SIREN}}-\mathrm{MAE}_{\mathrm{PINN}})/\mathrm{MAE}_{\mathrm{SIREN}}$.}
\scriptsize
\setlength{\tabcolsep}{3pt}
\renewcommand{\arraystretch}{1.1}
\resizebox{\linewidth}{!}{%
\begin{tabular}{lccccccccc}
\toprule
Sampling & $\mathcal{R}^{\mathrm{PINN}}_{\mathrm{equil}}$ (GPa/\textmu m) & $\mathcal{R}^{\mathrm{SIREN}}_{\mathrm{equil}}$ (GPa/\textmu m) & Ratio & $\mathcal{R}^{\mathrm{PINN}}_{\mathrm{compat}}$ (\textmu m$^{-2}$) & $\mathcal{R}^{\mathrm{SIREN}}_{\mathrm{compat}}$ (\textmu m$^{-2}$) & Ratio & MAE$_{\mathrm{PINN}}$ & MAE$_{\mathrm{SIREN}}$ & MAE red. (\%) \\
\midrule
1\%  & 110.07 & 201.77 & 1.83 & 15.22 & 33.60 & 2.21 & 0.02907 & 0.03227 &  9.91 \\
5\%  & 124.59 & 223.07 & 1.79 &  9.74 & 37.75 & 3.88 & 0.01990 & 0.01884 & -5.67 \\
10\% & 163.59 & 245.15 & 1.50 & 15.40 & 41.98 & 2.73 & 0.01702 & 0.01543 & -10.27 \\
25\% & 153.76 & 259.91 & 1.69 & 12.45 & 42.71 & 3.43 & 0.01545 & 0.01110 & -39.17 \\
50\% & 152.04 & 262.02 & 1.72 & 12.55 & 40.56 & 3.23 & 0.01483 & 0.00881 & -68.32 \\
75\% & 155.32 & 264.91 & 1.71 & 12.05 & 38.55 & 3.20 & 0.01464 & 0.00788 & -85.86 \\
\bottomrule
\end{tabular}%
}
\end{table}

\subsection*{Notes}

\begin{itemize}
\item The architectural requirements imposed by the elastic priors are consistently described as five: coordinate-based implicit \emph{ansatz}, non-vanishing second derivatives, frozen residual normalization, physics-weight ramp, and adaptive collocation refinement.
\item The angle channel $\theta$ is derived from the strain tensor rather than measured independently; the principal quantitative conclusions are therefore based on the three strain components, especially $\varepsilon_{xx}$.
\item The PINN improves physical self-consistency at every sampling fraction, while the data-only SIREN becomes more accurate once the retained sampling fraction exceeds the extreme-sparsity regime. This is consistent with the interpretation in the main text that the homogeneous isotropic equilibrium prior is useful as a sparse-data regulariser but becomes an approximation-induced bias when data are plentiful.
\end{itemize}

\end{document}